\documentclass [twocolumn,prl,amsmath,amssymb]{revtex4}

\usepackage [dvips]{graphicx}
\usepackage {dcolumn}
\usepackage {bm}

\begin {document}

\title {Pseudogap Value in the Energy Spectrum of LaOFeAs: Fixed Spin
Moment Treatment}

\author {M.A.~Korotin}
\affiliation {Institute of Metal Physics, Russian Academy of Sciences,
620041 Yekaterinburg GSP-170, Russia}

\author {S.V.~Streltsov}
\affiliation {Institute of Metal Physics, Russian Academy of Sciences,
620041 Yekaterinburg GSP-170, Russia}

\author {A.O.~Shorikov}
\affiliation {Institute of Metal Physics, Russian Academy of Sciences,
620041 Yekaterinburg GSP-170, Russia}

\author {V.I.~Anisimov}
\affiliation {Institute of Metal Physics, Russian Academy of Sciences,
620041 Yekaterinburg GSP-170, Russia}


\begin {abstract}

The experimental data available up to date in literature corresponding to
the paramagnetic - spin density wave transition in nonsuperconducting
LaOFeAs are discussed. In particular, we pay attention that upon spin
density wave transition there is a relative decrease of the density of
states on the Fermi level and a pseudogap formation. The values of these
quantities are not properly described in frames of the density functional
theory. The agreement of them with experimental estimations becomes more
accurate with the use of fixed spin moment procedure when iron spin moment
is set to experimental value. Strong electron correlations which are not
included into the present calculation scheme may lead both to the decrease
of spin moment and renormalization of energy spectrum in the vicinity of
the Fermi level for correct description of discussed characteristics.

\end {abstract}

\maketitle

Stimulated by the discovery of a new class of high-T$_c$ superconductors on
the base of LaOFeAs compound~\cite {Kamihara-08}, a lot of investigations
of the electronic and magnetic structure of this nonsuperconducting parent
compound in frames of the Density Functional Theory (DFT) were
performed~\cite {Singh-08, Cao-08, Mazin-08, Xu-08, Ma-08}. These 
calculations were successful in prediction of not only magnetic
instability~\cite {Singh-08, Xu-08}, but even an exact type of magnetic
structure of LaOFeAs~\cite {Mazin-08, Dong-08}. 

Calculated iron magnetic moment is close to 2~$\mu_B$~\cite {Ma-08,
Pickett-08}. However, experiments indicate much smaller value. Powder
neutron diffraction measurements~\cite {Cruz-08} give 0.36(5)~$\mu_B$.
Local probe measurements of magnetic properties of LaOFeAs such as
$^{57}$Fe M{\"o}ssbauer spectroscopy~\cite {Kitao-08} together with muon
spin relaxation~\cite {Klauss-08} indicate the values of $\sim$0.35~$\mu_B$
and 0.25(5)~$\mu_B$, correspondingly.

The situation when DFT calculations predict larger spin magnetic moment in
comparison with experimental one is rare and known only for few systems
(e.g. MnSi, ZrZn$_2$ etc.). The inconsistency between experimental and 
calculated magnetic moments in these materials may be ascribed to spin 
fluctuations which lead to the suppression of magnetic moment~\cite
{Mazin-03}. Nevertheless, LaOFeAs is an outstanding compound even among
these systems because the ratio $\mu_{calc}/\mu_{exp}$ is  extraordinary
large, $\sim$6, and more importantly since Fe ions in simple atomic picture
is expected to have $S=2$ which cannot be easily suppressed by any quantum
fluctuations.

The other known experimental parameters which can be compared with their
theoretical values are specific heat coefficient $\gamma$ related to the
density of states (DOS) on the Fermi level $N(E_F)$ as
$\gamma=\frac{\pi^2}{3}k_B^2N(E_F)$ and Pauli susceptibility 
$\chi=\mu_B^2N(E_F)$. Specific heat coefficient can be extracted from the
low-temperature behavior of the heat capacity. Unfortunately, this
parameter is ill defined experimentally, i.e. strongly depends on the
temperature range used in the fitting procedure. It was estimated by
different groups to be 3.7~mJ/(mol$\cdot$K$^2$)~\cite {Dong-08}, 
0.9~mJ/(mol$\cdot$K$^2$)~\cite {McGuire-08} and
0.69~mJ/(mol$\cdot$K$^2$)~\cite {Mu-08}. However, $\gamma$ obtained in
nonmagnetic DFT calculations~\cite {Singh-08, Xu-08} overestimates the
largest experimental value almost in two times. The results of magnetic
calculations for the real striped antiferromagnetic structure~\cite
{Pickett-08} improve the situation. They are close to the intermediate
experimental $\gamma$ value. However, one could consider this coincidence
as accidental since the electronic structure of this antiferromagnetic
solution corresponds to the large iron magnetic moment. Susceptibility
calculated in frames of nonmagnetic DFT is
8.5$\times$10$^{-5}$~emu/mol~\cite {Singh-08}. At the same time the flat
region of experimental susceptibility curve has the value of
$\sim$50$\times$10$^{-5}$~emu/mol~\cite {Kamihara-08, Nomura-08} -- 6 times
larger than calculated one.

There is an experimental indication of the formation of partial energy gap
(or pseudogap) around the Fermi level which removes parts of the DOS or few
bands from the Fermi energy at the phase transition from paramagnetic to
spin density wave (SDW) state. Direct experimental estimations of pseudogap
$E_{pg}$ is based on the results of reflectivity measurements~\cite
{Dong-08}. The pseudogap where the decrease of optical absorption spectra
is observed at different temperatures corresponding to paramagnetic and SDW
states is in the range of 150$\sim$350~cm$^{-1}$ (19$\sim$43~meV). The
value of the pseudogap due to SDW formation evaluated from thermal
transport experiments equals to 210~K \cite {McGuire-08} which corresponds
to 18~meV. Results of temperature-dependent angle-integrated laser
photoemission study for fluorine doped compound indicate $\sim$100~meV
pseudogap~\cite {Ishida-08}, whereas high-resolution photoemission
spectroscopy~\cite {Sato-08} gives the pseudogap value above T$_c$ of
15-20~meV.

The evaluation of the relative reduction of the DOS in the vicinity of the
Fermi energy due to formation of such pseudogap is more indirect procedure.
Based on the results of susceptibility measurements~\cite {Nomura-08} one
can deduce from the values of Pauli-like susceptibility curves at high and
low temperatures that the change in the $N(E_F)$ does not exceed 20\%. The
same value of $\sim$20\% one could extract from the conductivity curves of
Ref.~\onlinecite {Dong-08} keeping in mind that conductivity is
proportional to $N^2(E)$ in the proximity of $E_F$. Analysis of the specific
heat at various temperatures~\cite {McGuire-08} suggests 70\% reduction of
$N(E)$ around $E_F$ at paramagnetic - SDW transition.

Thus for LaOFeAs, one can assume that there is a decrease in DOS  near the
Fermi energy  of tens percent with paramagnetic - SDW transition which
corresponds to pseudogap formation of 20$\sim$100~meV. SDW state is
characterized by iron spin magnetic moment of 0.2$\sim$0.4~$\mu_B$,
specific heat coefficient of 1$\sim$4~mJ/(mol$\cdot$K$^2$) and
susceptibility coefficient of $\sim$50$\times$10$^{-5}$~emu/mol.
Conventional DFT calculations (both nonmagnetic and antiferromagnetic) fail
in the correct description of these quantities.

In the present paper we show that fixed spin moment DFT calculations
with the magnetic moment fixed at experimental value can significantly
improve an agreement with the experiments with regard to the values of
specific heat, pseudogap and relative decrease of the $N(E_F)$ value.

\begin {figure} 
\includegraphics [width=0.495\textwidth]{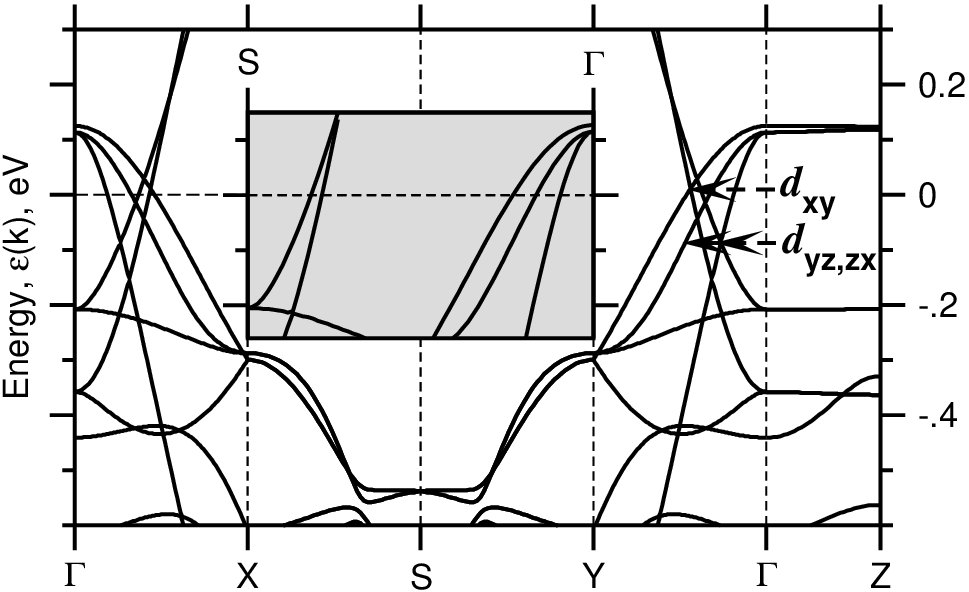}
\caption {\label {NM} Band structure of LaOFeAs obtained from nonmagnetic
calculation in BZ corresponding to enlarged ($\sqrt{2}a \times \sqrt{2}a
\times c$) unit cell. Grey insert: band structure along $S -\Gamma$ line
for conventional ($a \times a \times c$) unit cell. Fermi energy is zero.
For the bands crossing the Fermi level orbital projections are marked.}
\end {figure}

The calculations were performed within the framework of the Tight-Binding
Linear Muffin-Tin Orbitals (TB LMTO) method~\cite {Andersen-84} using the
Generalized Gradient Approximation (GGA), where exchange potential was
taken in Perdew-Wang form~\cite {Perdew-Wang}. Experimentally
determined~\cite {Cruz-08} structure parameters and atomic positions for
tetragonal phase and collinear striped antiferromagnetic order of Fe ions
in layer were used. We assumed ferromagnetic interlayer interaction due to
negligible influence of the antiparallel alignment of spins between
different FeAs layers and for the simplicity of discussion. The
La($6s$,$6p$,$5d$,$4f$), Fe($4s$,$4p$,$3d$), O($3s$,$2p$,$3d$) and 
As($4s$,$4p$,$4d$) orbitals were included into the basis set. The
integration in the course of the self-consistency iterations was performed
over a mesh of $18 \times 18 \times 12$ {\bf k}-points in the irreducible
part of Brillouin zone. We checked that such amount of {\bf k}-points is
enough for the precise calculation of the Fermi level position $E_F$ and
value of the density of states at the Fermi level $N(E_F)$. Fine mesh is
important due to the nesting bands near $E_F$. Calculations were performed
in  $\sqrt{2}a \times \sqrt{2}a \times c$ (four formula units) unit cell
appropriate for description of striped antiferromagnetic state.
Crystallographic $x$ and $y$ axes were directed from iron to its nearest
iron neighbors and ferromagnetic chains were running along $x$ direction. 

Fig.~\ref {NM} demonstrates the results of nonmagnetic calculations. The
band structure agrees with that obtained previously~\cite {Pickett-08}. 
The Fe bands mainly of $t_{2g}$ origin cross the Fermi level. One may
notice two-dimensional character of the band structure and clear signs of
Fermi surface nesting in $\Gamma-X$ and $Y-\Gamma$ directions. 

The nesting effect is usually illustrated in the figure of Fermi surface.
In order to reveal nesting in the simple band structure graph one may plot
it along $S-\Gamma$ line for conventional $a \times a \times c$ unit cell;
the result is given in the insert of Fig.~\ref {NM}. This $S-\Gamma$
direction corresponds to $Y-\Gamma$ (or $X-\Gamma$) line for enlarged
$\sqrt{2}a \times \sqrt{2}a \times c$ unit cell. When the unit cell is
doubled, left half of grey region folds to the right half. The cross of the
folded bands happens just on the Fermi level.

The specific heat and susceptibility coefficients recalculated from
$N(E_F)$ obtained in nonmagnetic DFT approach are $\gamma_{NM} =
5.3$~mJ/(mol$\cdot$K$^2$) and $\chi_{NM}$=7.2$\times$10$^{-5}$~emu/mol.
That agrees with values calculated before.

\begin {figure} 
\includegraphics [width=0.495\textwidth]{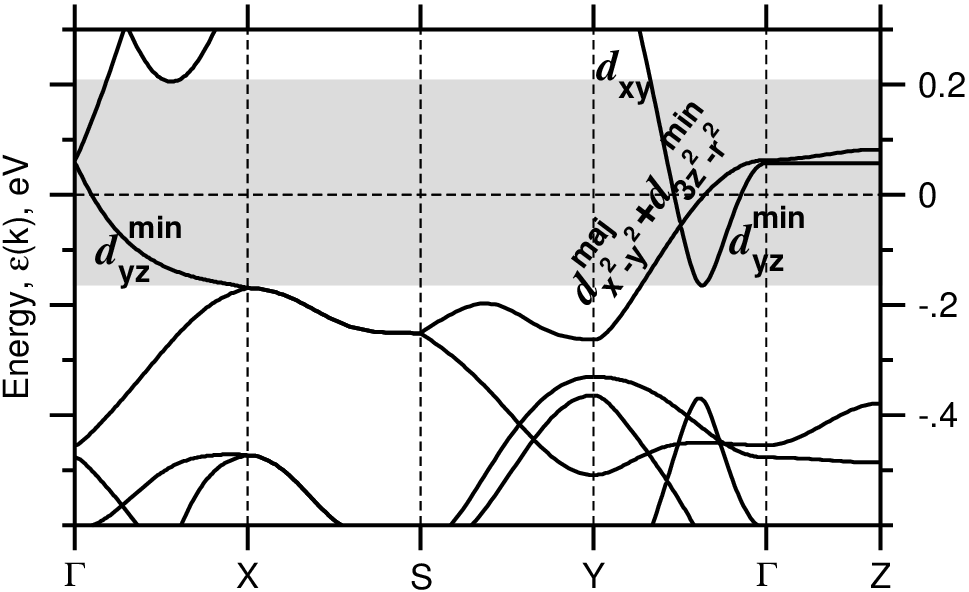}
\caption {\label {AFM} Band structure of LaOFeAs for striped
antiferromagnetic state. Pseudogap energy region is shown by a grey shadow
stripe. Fermi energy is zero. For the bands crossing the Fermi level
orbital projections are marked. Superscripts mean majority (maj) and
minority (min) spin projections. There are contributions from both majority
and minority states to the band $d_{xy}$.}
\end {figure}

The total energy difference between nonmagnetic and antiferromagnetic
(which is energetically more favored) states is 116~meV/(atom~Fe) which is
in a good agreement with the result of Ref.~\onlinecite {Pickett-08}.
Substantial energy gap between different magnetic solutions together with
the large magnetic moment (see below) shifts the system away  from the
quantum critical point, where spin fluctuations may play an important role.
Band structure for striped antiferromagnetic state is shown in Fig.~\ref
{AFM}. It differs essentially from the nonmagnetic picture. In particular,
in the vicinity of the Fermi level in $\Gamma - X$ direction there is only
one band of $d_{yz}^\downarrow$ character. The other 4 bands are moved away
from the Fermi level due to the Stoner splitting. In $Y - \Gamma$ direction
three bands remain. Two of them ($d_{xy}$ and $d_{yz}$) have the same
origin like as in nonmagnetic state. And third band originated from
$d_{x^2-y^2}^\uparrow$ and $d_{3z^2-r^2}^\downarrow$ orbitals which were
completely occupied in nonmagnetic case has been appeared around $E_F$.

One can define the pseudogap as an energy region around the Fermi level
where number of bands in Fig.~\ref {AFM} is essentially decreased in
comparison with that in Fig.~\ref {NM}. It is natural to define it between
maximum of parabola at $X$ {\bf k}-point and minimum of high-lying parabola
in $\Gamma - X$ direction (see grey stripe in Fig.~\ref {AFM}). The
pseudogap defined in such a way is estimated to be 380~meV which is much
larger than experimental expectations.

Calculated iron magnetic moment equals to 1.77~$\mu_B$ and specific heat
coefficient $\gamma_{MAG} = 0.99$~mJ/(mol$\cdot$K$^2$). This agrees with
the experimental estimations of $\gamma$~\cite {McGuire-08}. However, going
from nonmagnetic to antiferromagnetic phase the calculated value of
$N(E_F)$ changes by factor of 6 (600\% instead of tens percent). 

This inconsistency in the values of magnetic moment, width of the pseudogap
and too drastic change in $N(E_F)$ going from paramagnetic to magnetic
state demands the explanation. Below by the use of fixed spin moment
procedure we simulate experimental value of the magnetic moment and
investigate how $N(E_F)$ and the width of the pseudogap is changing upon
decrease of the spin moment value. 

\begin {figure}
\includegraphics [width=0.495\textwidth]{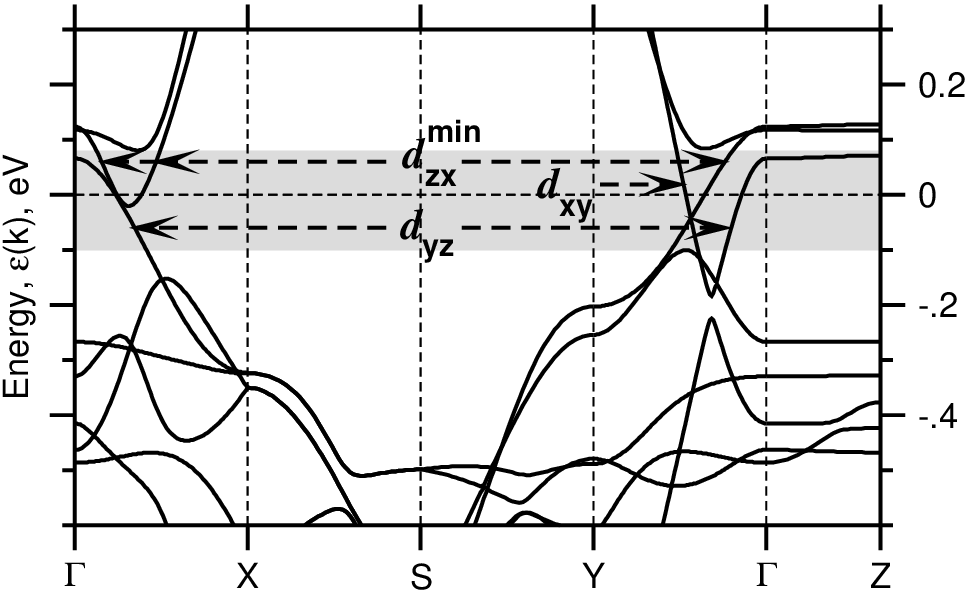}
\caption {\label {FSM-0.365} Band structure of LaOFeAs for striped
antiferromagnetic state with the fixed spin moment value 0.36~$\mu_B$. See
also Fig.~\ref {AFM} caption. }
\end {figure}

Band structure corresponding to the fixed spin moment value 0.36~$\mu_B$ is
presented in Fig.~\ref {FSM-0.365}. Now it looks very similar to
nonmagnetic picture. The remarkable difference occurs along $\Gamma - X$
direction: $d_{xy}$ and $d_{zx}$ bands are spin-splitted and first of them
is removed from the vicinity of the Fermi level whereas $d_{zx}^\downarrow$
is still crossing $E_F$. There is no dramatic reconstruction of bands along
$Y - \Gamma$ direction. One can conclude that upon transition from
nonmagnetic to magnetic state the first changes of band structure occur
along $\Gamma - X$ line; then the bands along $Y - \Gamma$ are involving to
the formation of magnetic moment.

Fixed spin moment calculation results in a significant growth of the
specific heat coefficient, which equals to $\gamma_{FSM} =
2.0$~mJ/(mol$\cdot$K$^2$) in this case in a good agreement with experiment.
The pseudogap which now could be defined as it's indicated by grey stripe
in Fig.~\ref {FSM-0.365} decreases down to 180~meV, which is still larger
than experimental one.

\begin {figure} 
\includegraphics [width=0.495\textwidth]{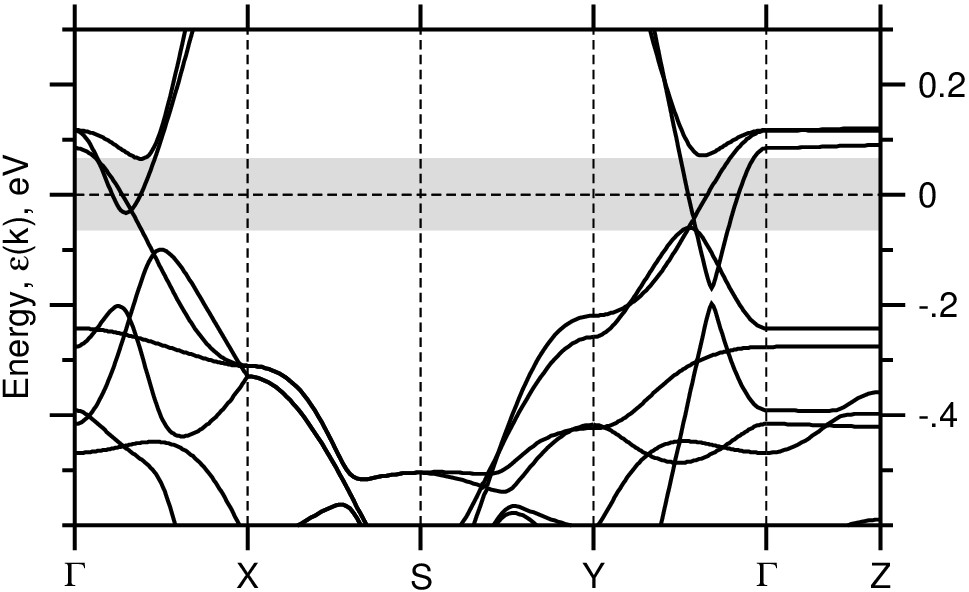}
\caption {\label {FSM-0.255} Band structure of LaOFeAs for striped
antiferromagnetic state with the fixed spin moment value 0.25~$\mu_B$. See
also Fig.~\ref {AFM} caption.}
\end {figure}

It is interesting to note that further decrease of the magnetic moment in
the fixed spin moment calculation (Fig.~\ref {FSM-0.255}) leads to even
better agreement with experiment with respect to the value of the
pseudogap. For one of the reported values of $\mu=0.25$~$\mu_B$ it
decreases down to 130~meV. The specific heat parameter is calculated to be
$2.4$~mJ/(mol$\cdot$K$^2$), and then the change in $N(E_F)$ in nonmagnetic
and magnetic states is only 55\% in reasonable agreement with experimental
estimations.

Semi-empirical fixed spin moment approach demonstrates that the
correspondence of experimentally known parameters of electronic structure
of LaOFeAs is essentially improved in comparison with conventional magnetic
calculations, if the magnetic moment is kept at low value $\sim$0.3~$\mu_B$
found in the experiment. However, even with reduced spin moment value the
calculated pseudogap (130~meV) remains larger than the experimental one
(20-40~meV).

In our opinion, an account for dynamical correlations which no doubts exist
for $d$ shell of iron may lead both to essential reduction of spin moment
value and renormalization of energy spectrum in the vicinity of the Fermi
level for decreasing of pseudogap.

This work is supported by Dynasty Foundation and International Center for 
Fundamental Physics in Moscow, Russian Foundation for Basic Research
through RFFI 07-02-91567 and 07-02-00041, Civil Research and Development
Foundation together with  Russian Ministry of science and education through
grant Y4-P-05-15, Russian president grant for young scientists
MK-1184.2007.2, grant of Ural division of RAS for young scientists.


\begin {thebibliography}{1}

\bibitem {Kamihara-08} Y. Kamihara, T. Watanabe, M. Hirano, and H. Hosono,
J. Am. Chem. Soc. {\bf 130}, 3296 (2008).

\bibitem {Singh-08} D.~J. Singh and M.~H. Du, arXiv: 0803.0429.

\bibitem {Cao-08} C. Cao, P.~J. Hirschfeld, and H.-P. Cheng, arXiv:
0803.3236.

\bibitem {Mazin-08} I.~I. Mazin, D.~J. Singh, M.~D. Johannes, and M.~H.
Du,  arXiv: 0803.2740.

\bibitem {Xu-08} G. Xu, W. Ming, Y. Yao, X. Dai, S. Zhang, and Z. Fang, 
arXiv: 0803.1282.

\bibitem {Ma-08} F. Ma and Z.-Y. Lu, arXiv: 0803.3286.

\bibitem {Dong-08} J. Dong, H.~J. Zhang, G. Xu, Z. Li, G. Li, W.~Z. Hu, D.
Wu,  G.~F. Chen, X. Dai, J.~L. Luo, Z. Fang, and N.~L. Wang,  arXiv:
0803.3426.

\bibitem {Pickett-08} Z.~P. Yin, S. Leb{\`e}gue, M.~J. Han, B. Neal,  S.~Y.
Savrasov, and W.~E. Pickett, arXiv: 0804.3355.

\bibitem {Cruz-08} C.~de~la Cruz, Q. Huang, J.~W. Lynn, J. Li,  W. Ratcliff
II, J.~L. Zarestky, H.~A. Mook, G.~F. Chen, J.~L. Luo,  N.~L. Wang, and P.
Dai, arXiv: 0804.0795.

\bibitem {Kitao-08} S. Kitao, Y. Kobayashi, S. Higashitaniguchi, M. Saito,
Y. Kamihara, M. Hirano, T. Mitsui, H. Hosono, and M. Seto, arXiv: 0805.0041.

\bibitem {Klauss-08} H.-H. Klauss, H. Luetkens, R. Klingeler, C. Hess,
F.~J. Litterst, M. Kraken, M.~M. Korshunov, I. Eremin, S.-L. Drechsler, R.
Khasanov, A. Amato, J. Hamann-Borrero, N. Leps, A. Kondrat, G. Behr, J.
Werner, and B. B{\"u}chner, arXiv: 0805.0264.

\bibitem {Mazin-03} I.~I. Mazin, D.~J. Singh, and A. Aguayo,  in
Proceedings of the NATO ARW on Physics of Spin in Solids: Materials,
Methods and Applications, edited by  S. Halilov (Kluwer, Dordrecht, 2003).

\bibitem {McGuire-08} M.~A. McGuire, A.~D. Christianson, A.~S. Sefat, R.
Jin, E.~A. Payzant, B.~C. Sales, M.~D. Lumsden, and D. Mandrus, 
arXiv: 0804.0796.

\bibitem {Mu-08} G. Mu, X. Zhu, L.Fang, L. Shan, C. Ren, and H.-H. Wen,
arXiv: 0803.0928.

\bibitem {Nomura-08} T. Nomura, S.~W. Kim,  Y. Kamihara, M. Hirano, P.~V.
Sushko, K. Kato, M. Takata,  A.~L. Shluger, and H. Hosono, arXiv: 0804.3569.

\bibitem {Ishida-08} Y. Ishida, T. Shimojima, K. Ishizaka, T. Kiss, M.
Okawa, T. Togashi, S. Watanabe, X.-Y. Wang, C.-T. Chen, Y. Kamihara, M.
Hirano, H. Hosono, and S. Shin, arXiv: 0805.2647.

\bibitem {Sato-08} T. Sato, S. Souma, K. Nakayama, K. Terashima, K.
Suguwara, T. Takahashi, Y. Kamihara, M. Hirano, and H. Hosono, arXiv:
0805.3001.

\bibitem {Andersen-84} O.~K. Andersen and O. Jepsen, Phys. Rev. Lett. {\bf
53}, 2571 (1984).

\bibitem {Perdew-Wang} J.~P. Perdew and Y. Wang, Phys. Rev. B {\bf 45},
13244 (1992).

\end {thebibliography}
\end {document}